\documentclass[preprint,11pt, aps,numerical,graphicx,superscriptaddress,longbibliography,showkeys,utf8,onecolumn]{revtex4-2}

\renewcommand{\thesection}{\arabic{section}}

\makeatletter
\renewcommand{\p@subsection}{}
\renewcommand{\p@subsubsection}{}
\makeatother

\usepackage[margin=1.5in]{geometry}
\usepackage{tabularx}
\usepackage{amssymb}
\usepackage[utf8]{inputenc}

\usepackage{amsmath}
\usepackage{amsfonts}

\usepackage{geometry}

\usepackage{caption}
\usepackage{natbib}
\usepackage{epstopdf}
\usepackage{graphicx}
\usepackage{subcaption}
\usepackage{todonotes}
\usepackage{tikz}
\usetikzlibrary{positioning}
\usepackage{float}

\usepackage{sistyle}
\usepackage[]{booktabs}
\usepackage[]{setspace}
\usepackage{hyperref}

\usepackage[english]{babel}

\usepackage{grffile}

\usepackage{chngcntr}

\usepackage{tikz}

\begin{document}
 \renewcommand\texteuro{FIXME} 
\def\SymbReg{\textsuperscript{\textregistered}}

\title{\bfseries The ROSMA-PTV concept and the impact of rotational setup uncertainties on the tumor control probability in canine brain tumors }

\author{Stephan Radonic}
\email[Author to whom correspondence should be addressed. Electronic mail: ]{stephan.radonic@uzh.ch}
\affiliation{Department of Physics, University of Zurich, Zurich, Switzerland}
\affiliation{Division of Radiation Oncology, Small Animal Department, Vetsuisse Faculty, University of Zurich, Zurich, Switzerland}

\author{Jürgen Besserer}
\affiliation{Radiotherapy Hirslanden AG, Rain 34 , Aarau, Switzerland}

\author{Carla Rohrer Bley}
\affiliation{Division of Radiation Oncology, Small Animal Department, Vetsuisse Faculty, University of Zurich, Zurich, Switzerland}

\author{Uwe Schneider}
\affiliation{Department of Physics, University of Zurich, Zurich, Switzerland}
\affiliation{Radiotherapy Hirslanden AG, Rain 34 , Aarau, Switzerland}

\author{Valeria Meier}
\affiliation{Department of Physics, University of Zurich, Zurich, Switzerland}
\affiliation{Division of Radiation Oncology, Small Animal Department, Vetsuisse Faculty, University of Zurich, Zurich, Switzerland}

\date{\today}

\begin{abstract}
	\textbf{Purpose:}   In this modelling study we pursued two main goals. The first was to establish a new CTV-to-PTV expansion which considers the closest and most critical organ at risk (OAR). The second goal was to investigate the impact of the planning target volume (PTV) margin size on the tumor control probability (TCP) dependent on the geometrical setup uncertainties. The aim was to achieve a smaller margin expansion close to the OAR while allowing a moderately larger expansion in less critical areas further away from the OAR and most importantly while maintaining the TCP.  \\
	\textbf{Methods and Materials:} Imaging data of radiation therapy plans from pet dogs which had undergone radiation therapy for brain tumor were used to estimate the clinic specific rotational setup uncertainties. A Monte-Carlo methodology using a voxel-based TCP model was used to quantify the implications of rotational setup uncertainties on the TCP. A combination of algorithms was utilized to establish a computational CTV-to-PTV expansion method based on probability density. This was achieved by choosing a center of rotation close to an OAR. All required software modules were developed and integrated into a software package which directly interacts with the Varian Eclipse treatment planning system.  \\
	\textbf{Results:}  The rotational setup uncertainties were obtained from 44 dog patients. Their frequency distributions were analysed and the mean values and standard deviations were determined. To investigate the impact on the TCP, several uniform and non-isotropic PTV targets were created. To ensure comparability and consistency standardized RT plans with equal optimization constraints were defined and automatically applied and calculated on these targets. The resulting TCPs were then evaluated and compared.     \\
	\textbf{Conclusion:} The non-isotropic margins were found to result in larger TCPs with smaller margin excess volume. Further we have presented an additional application of the newly established CTV-to-PTV expansion method for radiation therapy of the spinal axis of human patients.\\
\end{abstract}


\maketitle

\section{Introduction}

The success and efficacy of radiotherapy are in part determined by the dose distribution inflicted upon the tumor target. In classical fractionated radiotherapy (without radiosensitizer adjuvants, hypothermia, etc.) the spatio-temporal dose distribution is the only controllable parameter. The precision of delivery of the planned and calculated dose to a patient is heavily dependent on the tumor location and the equipment and techniques used. Those determine the precision limiting factors such as setup uncertainties, intra-fractional movement and inter-fractional positioning discrepancies. To ensure adequate dose coverage of the region to be treated, the clinical target volume (CTV) which encompasses the macroscopic and microscopic tumor burden, is further expanded to a planning target volume (PTV) to compensate the aforementioned uncertainties. The resulting PTV often lies within close proximity or even overlaps with organs at risk (OAR), thus research has concentrated on the reduction of the margins in order to reduce the risk of toxicity while maintaining tumor control. 

In principle all positional uncertainties could be corrected for in an ideal radiation therapy setting, using a linear accelerator and imager with perfect precision and a robotic couch. In such a scenario adding a margin to the CTV would be unnecessary. However, in realistic setups, especially without robotic couch an adequate margin remains crucial. The conventional approach to PTV definition is by uniform expansion of the CTV by a constant margin as proposed by ICRU \citep{10.1093/jicru/os26.1.Report50}. Numerous studies investigated the impact of geometric uncertainties on the tumor control and proposed margin recipes and the direct incorporation thereof into the treatment planning optimization instead of relying on the uniform PTV margin concept \citep{Stroom1999, VanHerk2000, VanHerk2002,RotheArnesen2008,Deveau2010,Chen2011,Jin2011,Witte2017,Miao2019,Selvaraj2013}. In particular \citet{Selvaraj2013} have investigated the impact of systematic and random geometric translational uncertainties on the tumor control depending on the margin size and fractionation. This was done for an artificial spherical CTV target with artificial spherical and four field brick dose distributions.

In this in-silico modelling study, our goals were to establish a new CTV-to-PTV expansion which considers the closest and most critical OAR, as well as to investigate as close to clinical reality as possible, the impact of the PTV margin size on the tumor control probability (TCP) dependent on the the geometrical setup uncertainties. The aim was to achieve a smaller margin expansion close to the OAR while allowing a moderately larger expansion in less critical areas further away from the OAR and most importantly while maintaining the TCP. A region of smallest margin (ROSMA) close to the most critical OAR was defined in an area where the PTV was desired to be smallest. A Monte-Carlo based non-isotropic PTV margin generation method, which is similar to the methodology proposed by \citet{Stroom1999} was devised. 

For quantification of the tumor control probability of the inhomogeneous dose distributions of the treatment plans we used a Monte-Carlo TCP calculation methodology which is reliant on the voxel-based TCP model proposed by \citet{Webb1993,Radonic2021}. The model was extended to incorporate inter-fractional temporal variations of the dose distribution. 
An IMRT plan template was applied, optimized and calculated for the different uniform and non-isotropic PTV targets. Subsequently, the resulting TCPs were evaluated and compared. The non-isotropic margins were found to result in larger TCPs with smaller margin excess volume.

\section{Methods and Materials}

In fractionated radiotherapy before each fraction the patient has to be positioned appropriately. For each fraction the positioning slightly varies. The variation magnitudes and the ability to correct them depends on the specific tumor site treated and on the equipment and procedures which are used. To account for the setup uncertainties, it is an established procedure to expand the CTV by a safety margin, which yields the PTV. The following section describes our methodology of examining the magnitude of the setup uncertainties occurring in our specific setup, its impact on the treatment outcome and the establishment of a CTV-to-PTV generation method.

\subsection{Setup, volumes and margin generation}
\subsubsection*{Patient inclusion criteria}

In this retrospective modelling study imaging data sets from pet dogs, which had undergone radiation therapy for brain tumor at the Vetsuisse Faculty of the University of Zurich in the period from 2007 to 2019 were used. Inclusion criteria were a simulation CT dataset as reference image and at least four cone-beam computed tomography (CBCT) datasets used for position verification during the course of radiation therapy. 

\subsubsection*{Positioning and verification}

All dogs underwent a short general anesthesia with endotracheal intubation using routine anesthetic protocols. They were positioned in a rigid positioning system consisting of a custom-made maxillary dental mold bite block (President The Original, Putty Soft, Coltène, Whaledent AG, Altstaetten, Switzerland) on a a polycarbonate tray that supported the maxilla and a vacuum cushion (BlueBag BodyFix, Elekta AB, Stockholm, Sweden) that supported the thorax and front legs. Position was verified with on-board imaging with kilovolt- (kV) CBCT using bone match (Clinac iX with OBI, Varian, Palo Alto, California, USA). In our setup, which uses a 4-degree of freedom couch, translational variations as well as the rotational yaw variation can be and are corrected for. However, roll and pitch variations can not be corrected for with the setup used. 

\subsubsection*{Definition of volumes}
\label{sec:delrosma}

To achieve a smaller margin expansion close to the OAR while allowing a moderately larger expansion in less critical areas further away from the OAR, a new structure called region of smallest margin (ROSMA) was created in the contouring workspace. The ROSMA is essentially a singular point with respect to which the daily CBCT to planning CT registrations is performed most carefully. For practical purposes however, a finite size structure is necessary.  The ROSMA was delineated depending on the most critical organ at risk (OAR) nearby. It can be the OAR itself or a part thereof.     
In case of the optic chiasm as most critical OAR, the optic chiasm structure was duplicated and used as ROSMA. In case of the brainstem, brain or eye as most critical OAR, the 3D brush tool with a 0.5cm diameter was used to draw a sphere. The latter was placed in the region where the PTV margin was desired to be smallest; i.e. between the clinical target volume (CTV) and the brainstem, brain or ocular bulb, respectively. In general the ROSMA has to be manually defined by the radiation oncologist as performed in the present study.

\subsubsection*{Registration of CBCTs to simulation CT}
\label{sec:cbctreg}

\begin{figure}[hbt]
\includegraphics[trim=50 120 200 10,width=0.7\columnwidth]{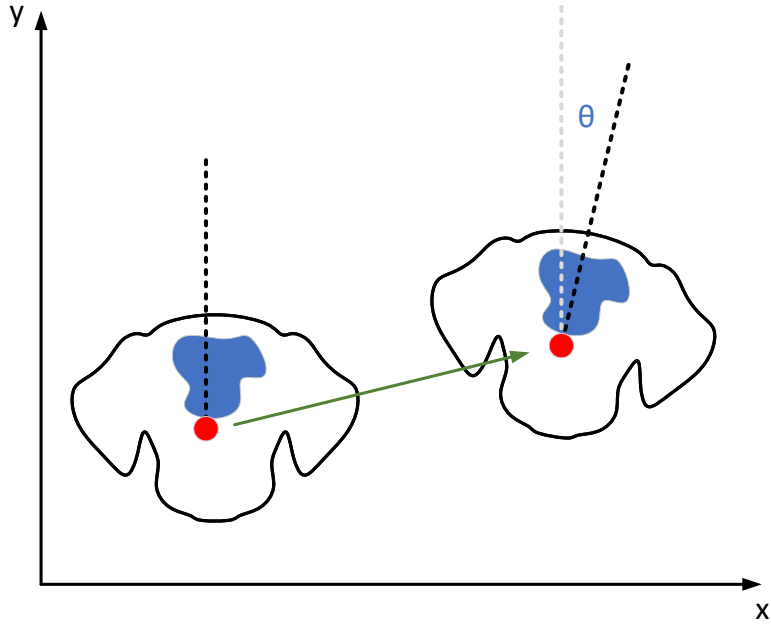}
\caption{Illustration of the rotational variation (roll angle) when the dog patients are positioned}
\label{fig:reg_illu}
\end{figure}

Bone registration of the skull was performed in the image registration workspace of the External Beam Planning system (Eclipse™ Planning system, Varian Oncology Systems, Palo Alto, California, USA). Each CBCT was retrospectively registered to the planning CT using the auto matching tool with bone intensity range from 200-1700 Hounsfield units and limiting the region of interest to the skull. Each registration corresponds to a transformation matrix $\mathbf{T}$~\eqref{eq:regmatrix}, which contains the rotations (roll $\phi$, pitch $\theta$, yaw $\psi$) and translations ($x_t$, $y_t$, $z_t$).
\begin{align}
\label{eq:regmatrix}
\mathbf{T}=
 \begin{pmatrix}
 \cos\psi\cos\theta & \cos\psi\sin\theta\sin\phi - \sin\psi\cos\phi & \cos\psi\sin\theta\cos\phi+\sin\psi\sin\phi & x_t \\
 \sin\psi\cos\theta & \sin\psi\sin\theta\sin\phi+\cos\psi\cos\phi & \sin\psi\sin\theta\cos\phi-\cos\psi\sin\phi & y_t \\
 -sin\theta & \cos\theta\sin\phi & \cos\theta\cos\phi & z_t \\
 0 & 0 & 0 & 1
 \end{pmatrix}
 \end{align}
The desired roll and pitch rotation angles were read out from the matrix. This is illustrated for the roll angle in Figure~\ref{fig:reg_illu}. The registrations were done for a total of 44 dogs and 220 CBCTs.

From the obtained set of roll and pitch variation angles, a Gaussian probability distribution was calculated for both rotation axis. The probability distributions were then used for Monte Carlo sampling in the PTV generation and the TCP estimation procedures as described in the respective sections.

\subsubsection*{Implementation}
All patients were planned using the Varian Eclipse™ Treatment Planning System (TPS), for which Varian provides an C\#/.NET based API called ESAPI (Eclipse Scripting Application Programmer Inferface). The API provides methods for data access and manipulation and allows the automatization of the various aspects of treatment planning. The computing intensive procedures were implemented in C++ and OpenMP was used to parallelize parts of the code. To interface the C++ code with the C\# code a wrapping layer was created using Microsoft C++/CLR. Google Protobufs was used to create the data interface for the three dimensional dose distribution and the region of interest (ROI) structures.

\subsubsection*{Voxelization}
\label{sec:voxel}
The contoured structures such as the CTV, PTV and OARs in DICOM format, as retrieved via the Varian ESAPI, are basically sets of slices. Each slice holds an ordered sequence of points (x,y,z - triplets) defining a contour (cite DICOM standard). For the required processing the CTV structures need to be voxelized. 
The grid resolution (voxel size) is chosen such that it matches the CT resolution, with which the required grid size is determined by the contour points with the minimal and maximal coordinates present in the structure. Accordingly, a mapping \eqref{eq:map}
\begin{equation}
m: (x,y,z) \rightarrow (i,j,k)
\label{eq:map}
\end{equation}
 \begin{equation}
 m^{-1} : (i,j,k) \rightarrow (x,y,z)
 \label{eq:invmap}
 \end{equation}
 between the points coordinates $\mathbf{r}=(x,y,z)^T \in\mathbb{R}$ and grid indices $\mathbf{IDX} =(i,j,k)^T \in\mathbb{Z^+}$ and vice versa \eqref{eq:invmap} is implied by \eqref{eq:coordstoidx} and \eqref{eq:idxtocoords}.
\begin{equation}
\mathbf{IDX} = m(x,y,z) = \lfloor (\mathbf{r}+\mathbf{S})/\mathbf{RES} \rfloor
\label{eq:coordstoidx}
\end{equation}
\begin{equation}
\mathbf{r} = m^{-1}(i,j,k) = \mathbf{RES}\odot\mathbf{IDX}-\mathbf{S}+\frac{\mathbf{RES}}{2}
\label{eq:idxtocoords}
\end{equation}
where $\mathbf{RES} = (x_{res}, y_{res}, z_{res})^T$ is the grid resolution and $\mathbf{S}$ is the coordinate transformation which translates all coordinates into positive domain and is given by  
\begin{equation}
\mathbf{S} = -\min \{x,y,z \in \text{CTV}\}
\end{equation} 
For all slices, iterating through the ordered sequence of contour points and using the \emph{Bresenham's line algorithm} \citep{10.1147sj.41.0025} the values lying on the contour polygon are set to one. Subsequently using the even-odd rule \cite{franklin}, the value of each voxel which lies inside the boundaries of the contour polygons, is set to one. Which-with the voxelization process is completed, all voxels belonging to the structure now have the integer value one, all other voxels have the value zero. 
\subsubsection*{Generation of the dynamic margin}
\label{sec:rosmaptv}
\begin{figure*}[ht]
\centering
\includegraphics[clip, trim=0.5cm 4.0cm 0.2cm 2.0cm, width=1.0\textwidth]{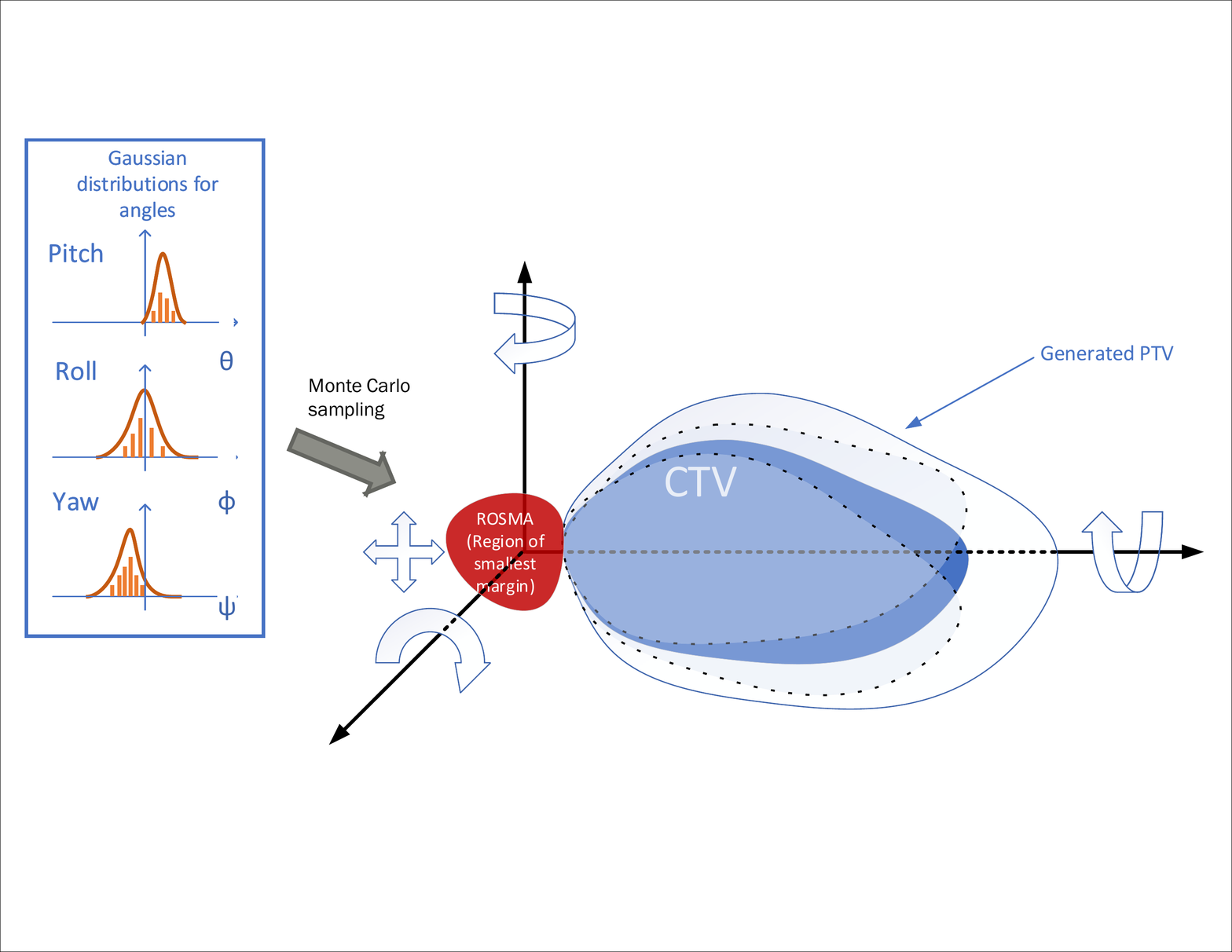}
\caption{Illustration of the Monte-Carlo based PTV generation}
\label{fig:rosmaptv}
\end{figure*}

First a stationary global three-dimensional voxel grid (integer array is initialized. 
As illustrated in Fig.~\ref{fig:rosmaptv} in a Monte Carlo procedure the rotation angles are sampled from the probability distributions derived from the CBCT registrations. Using the voxelized CTV structure obtained in the voxelization procedure, we iterate through all voxels. Using the sampled rotation angles $(\theta,\psi,\phi)$ a rotational transformation is applied to the midpoint coordinates of each voxel $W_{\mathbf{IDX}}$, rotating it around the center of mass of the delineated ROSMA $\mathbf{r_{ROSMA}}$ 
\begin{equation}
\mathbf{r'}=\mathbf{T^{-1}(r,r_{ROSMA})}\mathbf{R}(\theta,\psi,\phi)\mathbf{T(r,r_{ROSMA})}\cdot\mathbf{r}
\end{equation}
where $\mathbf{T(r,r_{ROSMA})}$ is the translation to the center of mass of the ROSMA, and $\mathbf{R}(\theta,\psi,\phi)$ is the rotation matrix.
The resulting coordinates $r'$ are mapped to the indices of the global voxel grid $\mathbf{IDX}_{\text{GLOB}}$ as in Eq.~\eqref{eq:coordstoidx}. We add the value of the rotated voxel to the corresponding voxel in the global grid. 
\begin{equation}
W^{\text{GLOB}}_{\mathbf{IDX_{\text{GLOB}}}} \mathrel{+}= W_{\mathbf{IDX}}
\end{equation} 
Thus after repeating the procedure for N times, the global voxel grid contains an occupancy probability distribution. The voxel values indicate how many times the particular voxel was inside the rotated CTV. An occurrence threshold can then be defined and subtracted from all voxels of the global grid. Which-after the voxels with a value of zero or above define our newly generated PTV. 
\begin{figure}[H]
\begin{tikzpicture}[scale=.8,every node/.style={minimum size=1cm},on grid]
   \begin{scope}[
           yshift=-83,every node/.append style={
           yslant=0.5,xslant=-1},yslant=0.5,xslant=-1
           ]
       \draw[step=4mm, black] (0,0) grid (5,5); 
       \draw[black,thick] (0,0) rectangle (5,5);
       \fill[black] (2.05,2.05) rectangle (2.35,2.35); 
       \fill[black] (1.65,2.05) rectangle (1.95,2.35); 
       \fill[black] (2.45,2.05) rectangle (2.75,2.35); 
       \fill[black] (2.05,2.45) rectangle (2.35,2.75); 
       \fill[black] (2.05,1.95) rectangle (2.35,1.65); 
	   \fill[blue] (1.65,2.85) rectangle (1.95,3.15); 
	   \fill[blue] (1.25,2.85) rectangle (1.55,3.15); 
	   \fill[blue] (2.05,2.85) rectangle (2.35,3.15); 
	   \fill[blue] (2.45,2.85) rectangle (2.75,3.15); 
	   \fill[blue] (2.85,2.85) rectangle (3.15,3.15); 
	   \fill[blue] (2.85,2.45) rectangle (3.15,2.75); 
	   \fill[blue] (2.85,2.05) rectangle (3.15,2.35); 
	   \fill[blue] (2.85,1.65) rectangle (3.15,1.95); 
	   \fill[blue] (2.85,1.25) rectangle (3.15,1.55); 
	   \fill[blue] (2.45,1.25) rectangle (2.75,1.55); 
	   \fill[blue] (2.05,1.25) rectangle (2.35,1.55); 
	   \fill[blue] (1.65,1.25) rectangle (1.95,1.55); 
	   \fill[blue] (1.25,1.25) rectangle (1.55,1.55); 
	   \fill[blue] (1.25,1.65) rectangle (1.55,1.95); 
	   \fill[blue] (1.25,2.05) rectangle (1.55,2.35); 
	   \fill[blue] (1.25,2.45) rectangle (1.55,2.75); 
       \fill[black] (1.65,2.45) rectangle (1.95,2.75); 
       \fill[black] (2.45,2.45) rectangle (2.75,2.75); 
       \fill[black] (2.75,1.95) rectangle (2.45,1.65); 
       \fill[black] (1.65,1.95) rectangle (1.95,1.65); 
      
   \end{scope}
   \begin{scope}[
           yshift=0,every node/.append style={
           yslant=0.5,xslant=-1},yslant=0.5,xslant=-1
           ]
       \fill[white,fill opacity=0.9] (0,0) rectangle (5,5);
       \draw[step=4mm, black] (0,0) grid (5,5); 
       \draw[black,thick] (0,0) rectangle (5,5);
       \fill[black] (2.05,2.05) rectangle (2.35,2.35); 
       \fill[black]   (1.65,2.05) rectangle (1.95,2.35); 
       \fill[black] (2.45,2.05) rectangle (2.75,2.35); 
       \fill[blue] (2.85,2.05) rectangle (3.15,2.35); 
       \fill[black] (2.05,2.45) rectangle (2.35,2.75); 
       \fill[blue] (2.45,2.85) rectangle (2.75,3.15); 
       \fill[blue] (2.05,2.85) rectangle (2.35,3.15); 
       \fill[blue] (1.65,2.85) rectangle (1.95,3.15); 
       \fill[blue] (1.25,2.85) rectangle (1.55,3.15); 
       \fill[blue] (0.85,2.85) rectangle (1.15,3.15); 
       \fill[blue] (1.25,2.45) rectangle (1.55,2.75); 
       \fill[blue] (1.25,2.05) rectangle (1.55,2.35); 
       \fill[blue] (1.25,1.65) rectangle (1.55,1.95); 
       \fill[black] (1.25,0.85) rectangle (1.55,1.15); 
     
       \fill[blue] (2.05,1.95) rectangle (2.35,1.65); 
       \fill[black] (1.65,2.45) rectangle (1.95,2.75); 
       \fill[blue] (2.45,2.45) rectangle (2.75,2.75); 
       \fill[blue] (2.85,2.85) rectangle (3.15,3.15); 
       \fill[blue] (2.75,1.95) rectangle (2.45,1.65); 
       \fill[blue] (1.65,1.55) rectangle (1.95,1.25); 
       \fill[blue] (1.65,1.15) rectangle (1.95,0.85); 
       \fill[blue] (1.65,1.15) rectangle (1.95,0.85); 
       \fill[black] (1.65,1.95) rectangle (1.95,1.65); 
       \fill[black] (1.25,1.55) rectangle (1.55,1.25); 
       \fill[blue] (0.85,1.55) rectangle (1.15,1.25); 
       \fill[blue] (0.85,1.15) rectangle (1.15,0.85); 
       \fill[blue] (1.25,0.75) rectangle (1.55,0.45); 
   \end{scope}
%
%
   \draw[-latex,thick,blue](-3,5)node[left]{ }
       to[out=0,in=90] (-.4,1.4);
   \draw[-latex,thick,blue](-3,5)node[left]{ }
       to[out=0,in=90] (0.8,1.15);
   \draw[-latex,thick,blue](-3,5)node[left]{blue: edge voxels with $W\geq0$}
       to[out=0,in=90] (0,2.8);
   \draw[-latex,thick,black](-3,-2)node[left]{black: voxels with $W\geq0$}
       to[out=0,in=200] (-1,-.9);
   \draw[thick,gray!70!black](6,4) node {slice k};
   \draw[thick,gray!70!black](6,-2) node {slice k-1};
\end{tikzpicture}
\caption{Illustration of contour extraction process}
\label{fig:contour}
\end{figure}
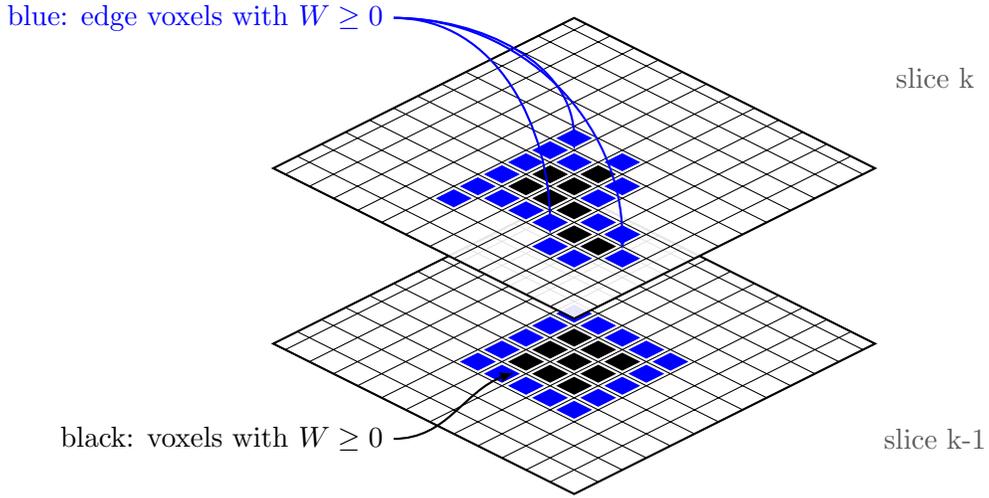

To write the generated PTV structure back to the TPS, the corresponding contour points have to be extracted from the voxel representation. For each slice $k$ iterating through all voxels $W_{i,j}$, the outermost voxels where $W_{i,j}\geq 0$ are found. The midpoints of those voxels define a contour. (see Figure \ref{fig:contour}) However the points are not yet ordered in a sequence as required by the TPS.
To create an ordered polygon from the unordered set of point a concave hulling algorithm from \citep{adaszewski-2019,Park2012ANC} is used. The obtained polygons are written back into the TPS as structure $PTV_{\text{GEN}}'$ and subsequently a boolean $OR$ operation between the segment volumes of the generated structure and the CTV is performed such that it is guaranteed for the CTV to be fully included in $PTV_{\text{GEN}}$ 
\begin{equation*}
CTV \subset PTV_{\text{GEN}}
\end{equation*}
as it is possible that for a very small threshold some CTV voxel are outside of the generated PTV margin.
That concludes the generation procedure of the PTV targets. An additional application of the generation procedure for humans is presented in the appendix.

\subsection{Treatment planning}
\label{sec:planning}

An nine field equispaced coplanar IMRT plan was devised. A minimal set of optimization criteria was chosen such that an homogeneous coverage of the designated target (PTV) and optimal conformity are ensured. Utilizing ESAPI, a script was created which automatically adjusts the devised plan template to each designated target (PTV) and subsequently runs the plan optimization, dose calculation and normalization. For the plan optimization the Varian Photon Optimizer Version 15.06.04 was used. The calculation model used was the Varian Anisotropic Analytical Algorithm Version 15.06.04. The dose calculation grid precision was set to one millimetre, which was the maximal precision available. For plan normalization a helper structure in the center of the CTV was defined by subtracting an uniform margin $r_m=5\cdot\sqrt[3]{ \frac{3}{4\pi}V_{\text{CTV}}}$  from the CTV ($V_{\text{CTV}}$ is the CTV volume). The plan was then normalized such that the median dose inside the devised helper structure equals the prescribed dose. \\
\\
During the procedure we have noticed that repeated optimization runs of the same plan with all the same optimization criteria to a just minimally different target leads to quite different spatial dose distributions. To ensure comparability of the different plans, we have done the plan optimizations of the different targets iteratively. In a first step the plan was optimized and calculated for the smallest designated target $T_0$. Next the plan was copied and adjusted to the next larger target $T_0 \rightarrow T_1$. Then the optimization was run but not from scratch but rather starting from the immediate dose already calculated. The iteration $T_n\rightarrow T_{n+1}$ was continued up until the largest target. For the anisotropic margin targets, the plan of the uniform margin target, with the most similar volume to the designated anisotropic target, was used as the starting point.

\subsection{Estimation of tumor control probability}
\label{sec:tcpsim}
\begin{figure*}[ht]
\centering
\includegraphics[clip, trim=0.5cm 4.0cm 0.2cm 2.0cm, width=1.0\textwidth]{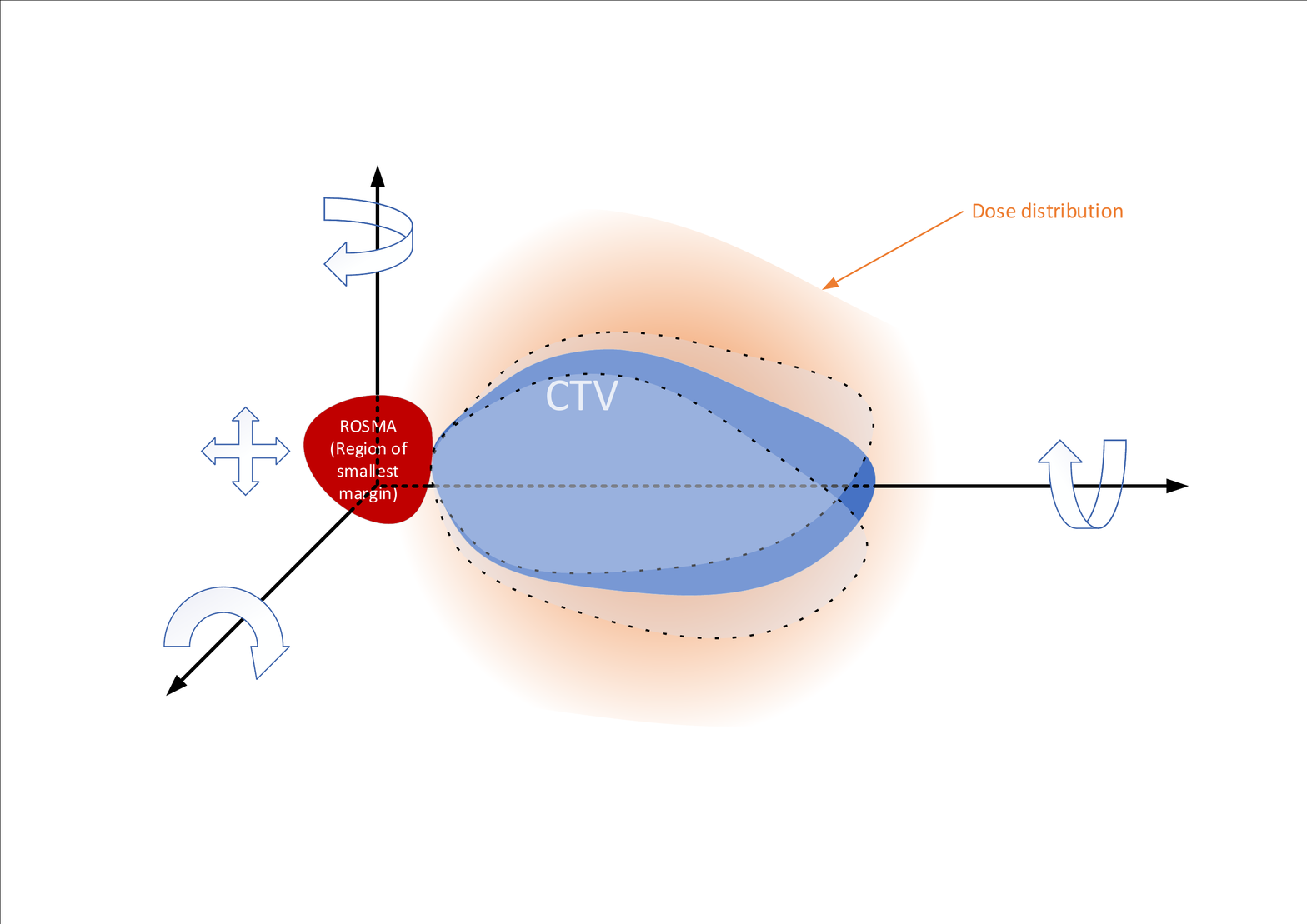}
\caption{Illustration of the Monte-Carlo based TCP estimation}
\label{fig:tcpmc}
\end{figure*}

Fractionated radiotherapy is simulated in a Monte-Carlo procedure. The TCP as modelled by \citet{10.1007978-3-642-48681-4_71} is given by

\begin{equation}
TCP = e^{-N_S}
\end{equation}
where $N_S$ is the number of surviving clonogenic cells. \citet{Webb1993} have derived a model for the TCP with non-uniform clonogenic cell density and non-uniform dose. 
We have extended the model to incorporate fractions and the temporal variations of the dose distributions, which are thereby implicated. The extended model is used in a Monte-Carlo procedure.%
Rotation angles are sampled from probability distributions obtained from CBCT registrations, as described in section \ref{sec:cbctreg}. 
As illustrated in Figure \ref{fig:tcpmc}, the voxelized CTV is rotated around the center of mass of the delineated ROSMA inside the dose distribution of the RT Plan.  Each rotation corresponds to a treatment fraction. For each voxel of the rotated CTV the dose exposure for the current fraction is recorded. Therewith the characteristic of the fractionated radiotherapy is modelled, that due to geometric positioning uncertainties, a particular CTV voxel gets a different dose at each fraction.

The recorded dose values are used to calculate the cell survival of the particular CTV voxel. The cell survival at fraction $f$ of a voxel $i$ is given by
\begin{equation}
S_i^f = S(D_i^f,\alpha,\beta)
\end{equation}
where $S$ is the survival model function and $D_i^f$ is the dose [Gy] which the voxel $i$ is exposed to at fraction $f$. In case of LQ Model \citep{fowlerjf} the cell survival function is given by
\begin{equation}
S(D_i^f,\alpha,\beta)=e^{-\alpha D_i^f -\beta (D_i^f)^2}
\end{equation}
In \citep{doi:10.12590007-1285-62-735-241} the LQ-model is combined with an assumed tumor repopulation factor. Further in \citep{Qi2006} the dependence of the survival rate on the elapse time is characterized by an exponential decrease.
The cell survival in voxel $i$ after $M$ fractions is 
\begin{equation}
S_i = \prod_{f=0}^M S_i^f
\end{equation}
The patient survival (TCP) after a follow-up period $\tau$ is then given by
\begin{equation}
TCP = \prod_{i=0}^N e^{-\rho_i V_i S_i e^{\gamma T}e^{a\tau}}
\end{equation}
where $e^{\gamma T}$ accounts for the effective tumor-cell repopulation rate and $e^{a\tau}$ characterizes exponential dependence of the survival rate on the elapse time. The radio-biological parameters for animal brain tumors, used for the TCP calculation, obtained from \cite{Radonic2021}, were $\alpha=0.36$, $\alpha/\beta = 8$, $a=0.9$ yr$^{-1}$, $T_D=5.0$ d. The TCPs were calculated for a follow-up time of $2$ years. \\
\\
In a Monte Carlo procedure the calculation is repeated $K$ times, this simulates a fictitious patient population undergoing the exact same radiation treatment protocol for the exact same tumor. For each simulation run the resulting TCP value is recorded. This yields a TCP frequency distribution. \\
\\ 
We integrated the frequency distribution and created a cumulative histogram. It represents the percentage of the fictitious simulated patient population having a TCP greater than or equal to the value in the corresponding TCP bin. We have chosen to present the  $TCP_{95\%}$, which means that in $95\%$ of the simulated treatments, the realised TCP was at least equal to the specified $TCP_{95\%}$ value.


\section{Results}

\subsection{Setup uncertainties}
\begin{figure}
	\begin{subfigure}[b]{\textwidth}
         \centering
         \includegraphics[width=\textwidth]{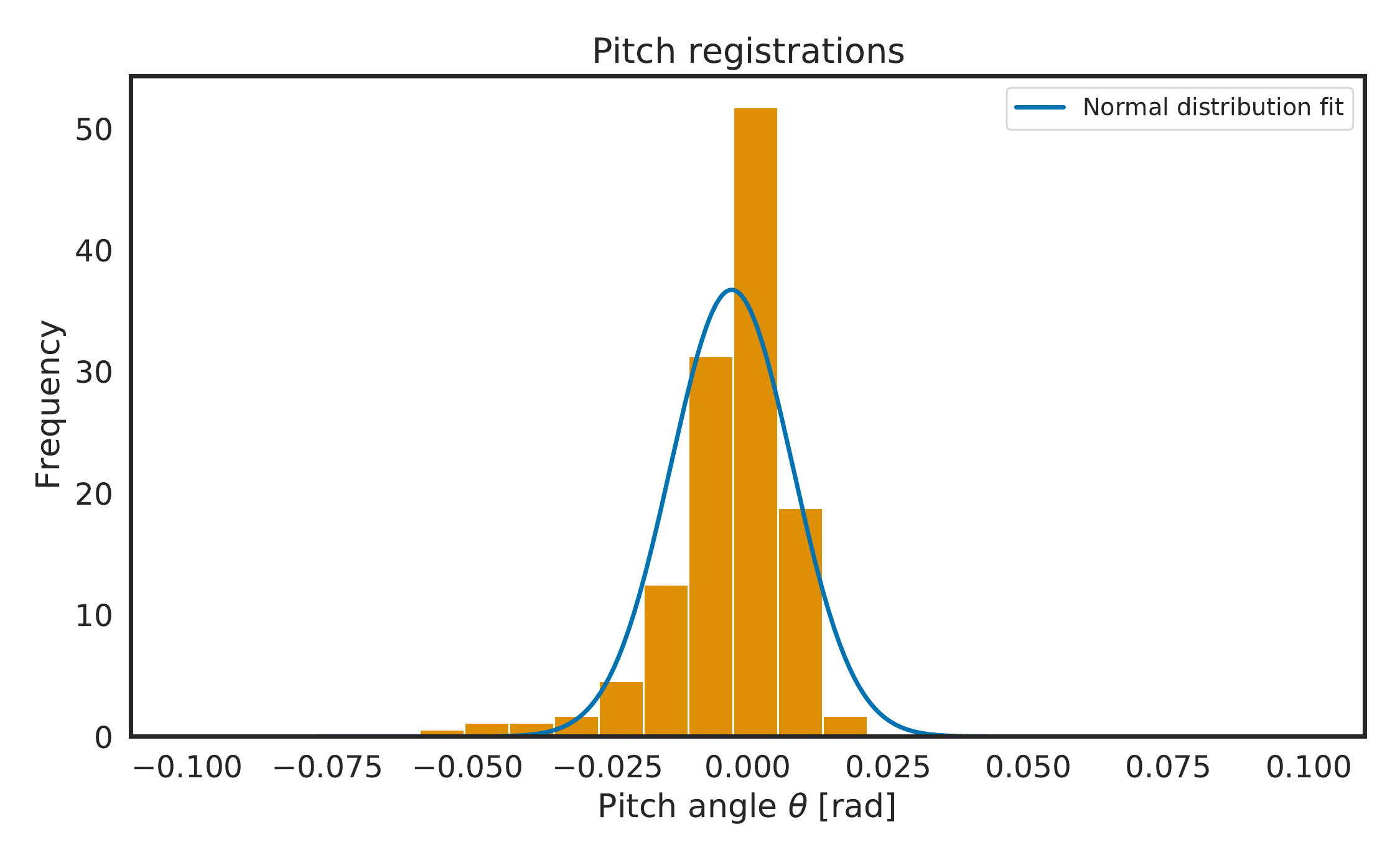}
         \caption*{}
         \label{fig:pitch_hist}
     \end{subfigure}
     \begin{subfigure}[b]{\textwidth}
         \centering
         \includegraphics[width=\textwidth]{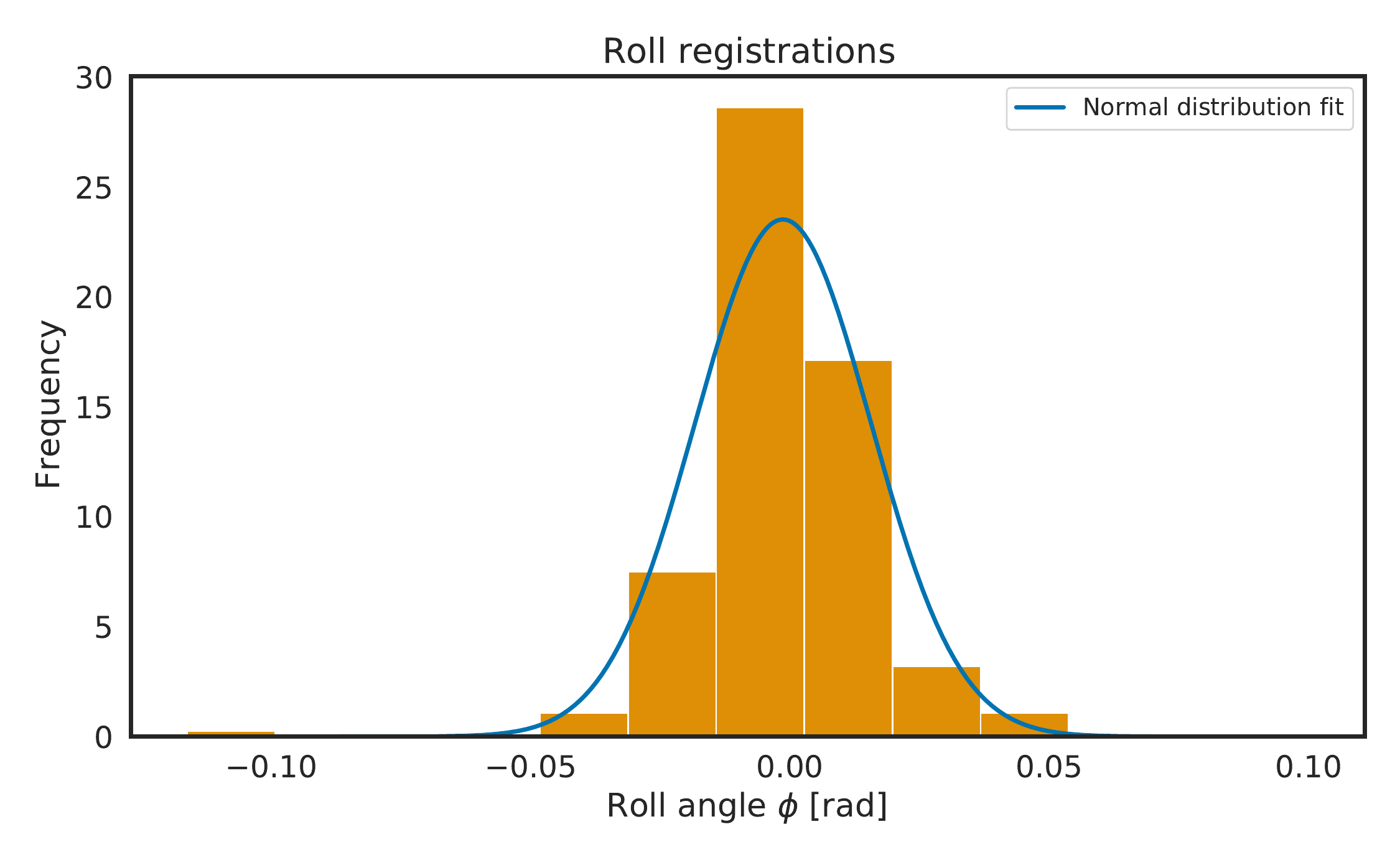}
         \caption*{}
         \label{fig:roll_hist}
     \end{subfigure}
\caption{The plots show the histograms of the rotations detected in registrations of CBCTs and corresponding fits of normal distributions; A: Roll angles; B: Pitch angles}
\label{fig:histogram_angles}
\end{figure}

In Figure~\ref{fig:histogram_angles} the roll and pitch angles obtained from registrations of positioning CBCTs (as explained in section~\ref{sec:cbctreg}) are plotted in histograms. The means of the pitch and roll angles were found to be $\mu_\theta=-0.003$ rad and $\mu_\phi=-0.001$ rad while the standard deviations were $\sigma_\theta=0.011$ rad and $\sigma_\phi=0.017$ rad.

\subsection{Impact of rotational uncertanties on TCP}

\begin{figure}
	\begin{subfigure}[b]{0.5\textwidth}
         \centering
         \includegraphics[height=0.4\textheight]{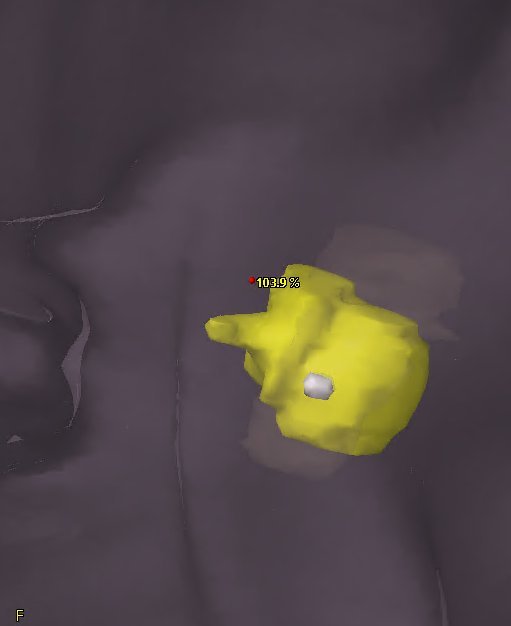}
         \caption*{A}
         \label{fig:org_ctv_3d}
     \end{subfigure}%
     \begin{subfigure}[b]{0.5\textwidth}
         \centering
         \includegraphics[height=0.4\textheight]{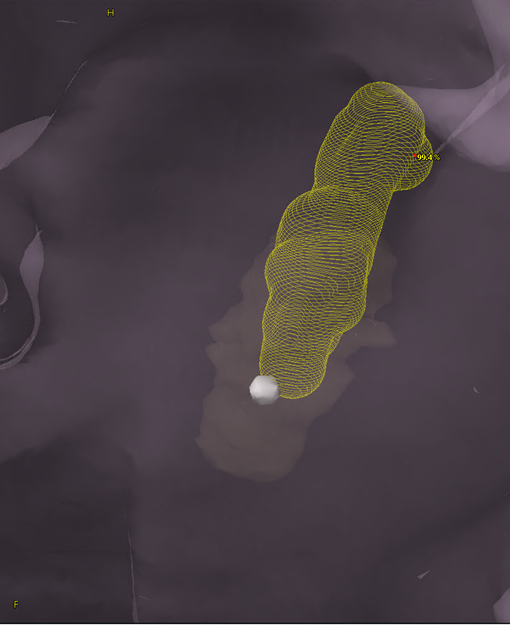}
         \caption*{B}
         \label{fig:art_ctv_3d}
     \end{subfigure}
\caption{The plots show the original and the artificial CTV (yellow) and the ROSMA ROI structures(grey). The ROSMA is the small spherical structure visible in the foreground; A: original CTV: irregular ellipsoid-like shape, B: artificial CTV: irregular cylindrical shape}
\label{fig:3d_ctv_rosma}
\end{figure}

\begin{figure}
	\begin{subfigure}[b]{\textwidth}
         \centering
         \includegraphics[width=\textwidth]{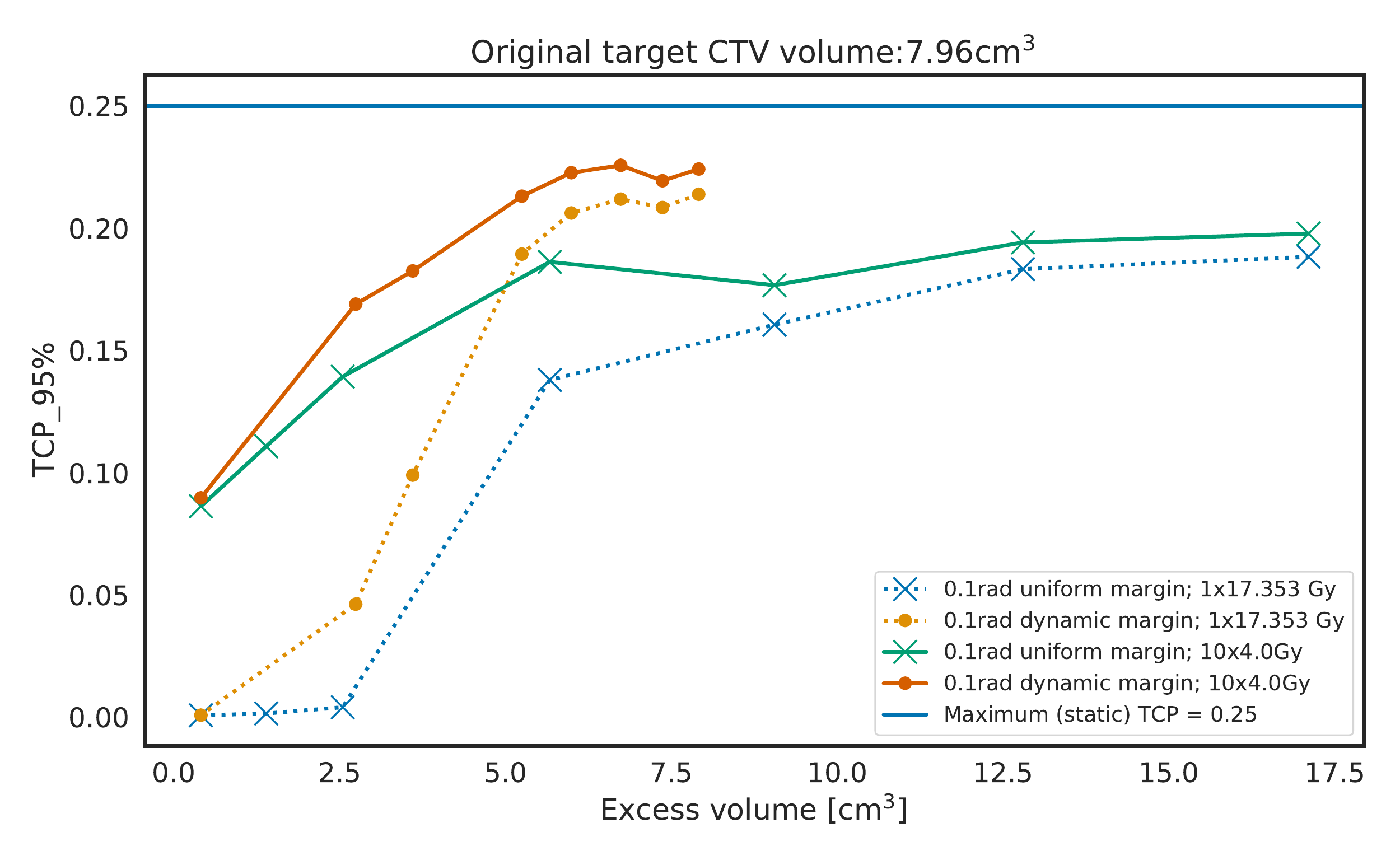}
         \caption*{A}
         \label{fig:org_ctv_0.1rad}
     \end{subfigure}
     \begin{subfigure}[b]{\textwidth}
         \centering
         \includegraphics[width=\textwidth]{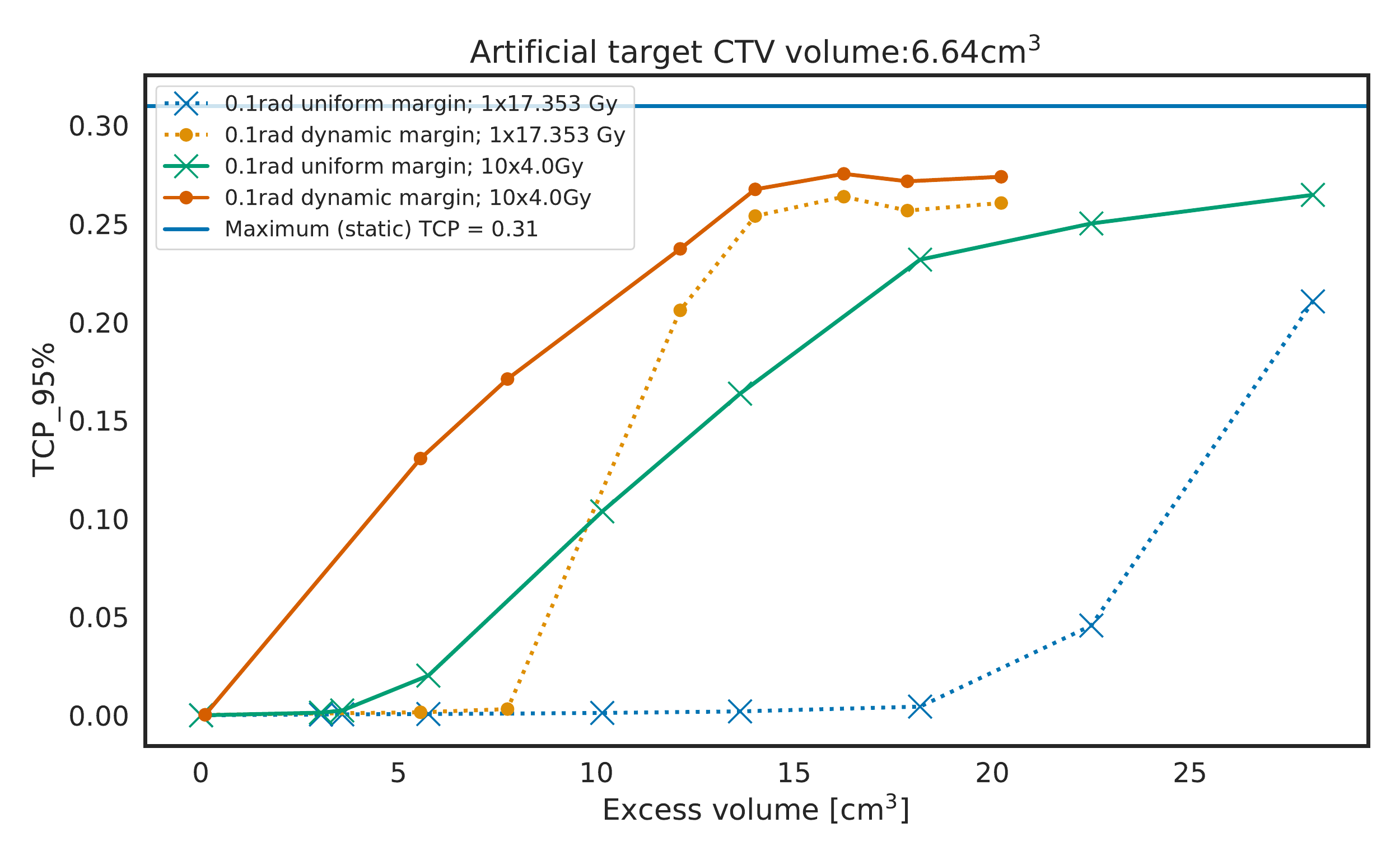}
         \caption*{B}
         \label{fig:art_ctv_0.1rad}
     \end{subfigure}
\caption{The plots show the comparison of the $TCP_{95\%}$ plotted versus the excess volume for the different uniform and anisotropic margin targets for the simulated single fraction and ten fractions treatments with assumed normally distributed rotational shifts with $\mu_\theta=\mu_\phi=0$ rad and $\sigma_\theta=\sigma_\phi=0.1$ rad; A: original CTV; B: artificial CTV}
\label{fig:tcp_vol_comp}
\end{figure}

\begin{figure}
 \centering
 \includegraphics[width=\textwidth]{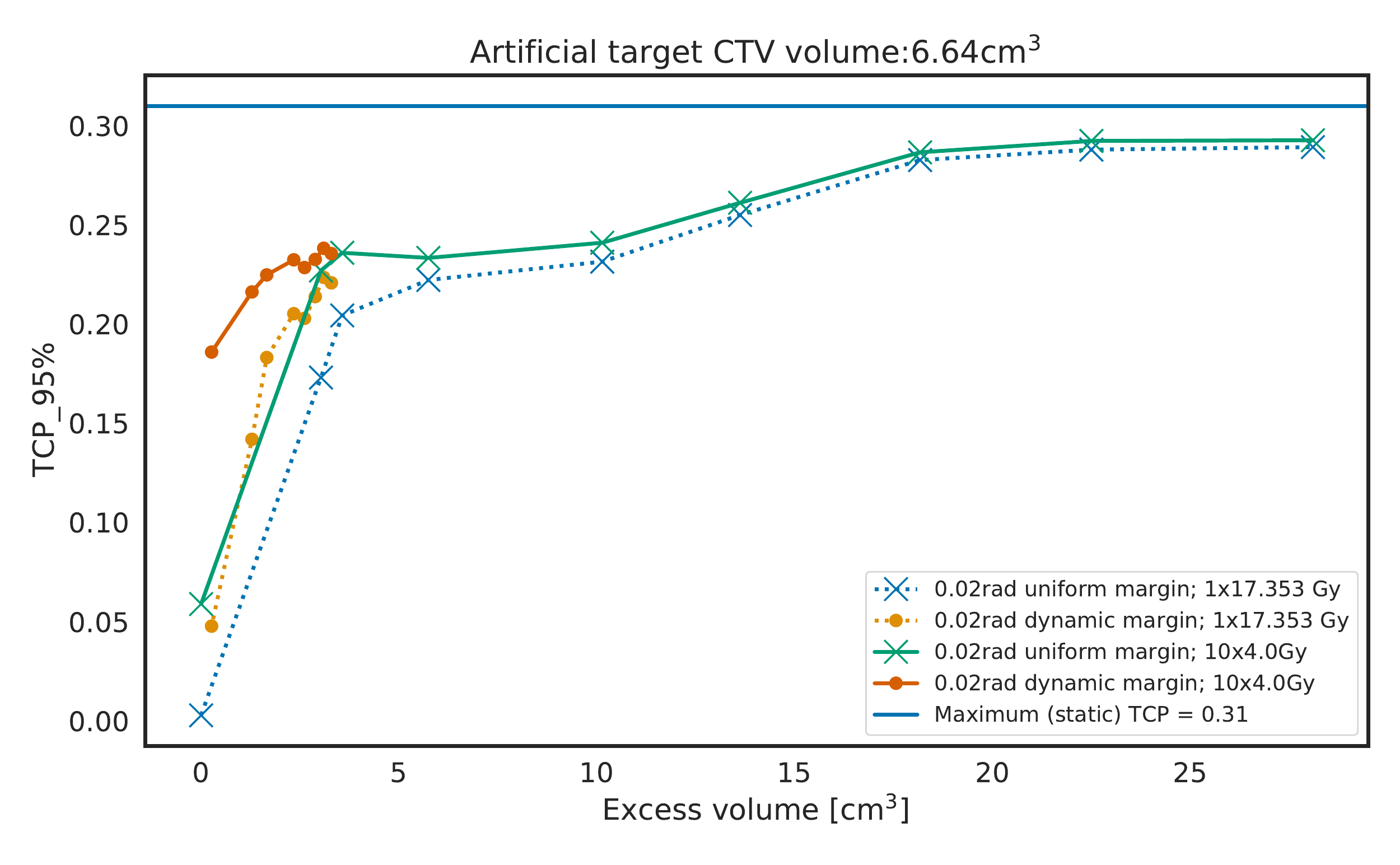}
         \caption{The plots show the comparison of the $TCP_{95\%}$ plotted versus the excess volume for the different uniform and anisotropic margin targets for the simulated single fraction and ten fractions treatments with assumed normally distributed rotational shifts with $\mu_\theta=\mu_\phi=0$ rad and $\sigma_\theta=\sigma_\phi=0.02$ rad for the artificial CTV}
         \label{fig:art_ctv_0.02rad}
\end{figure}

We investigated the impact of rotational uncertainties on the TCP depending on the margin size, for uniform margins as well as for anisotropic margins. For this purpose two CTVs with different geometrical shapes were selected. The first CTV was the original brain tumor CTV which was actually irradiated. The original CTV fairly resembles an ellipsoid with some distortions. With the intent to investigate the impact of the geometrical shape, a second CTV with similar volume was artificially drawn in the same region as the original CTV, to resemble a more cylindrical longitudinal shape. The CTVs and the ROSMA are depicted in Figure~\ref{fig:3d_ctv_rosma}.

The rotational roll and pitch uncertainties investigated were assumed to follow normal distributions $\mathcal{N}(\mu,\sigma^2)$ around 
$\mu=0$. We have considered uncertainty magnitudes with $\sigma_\theta=\sigma_\phi=0.1$, as well as $\sigma_\theta=\sigma_\phi=0.02$ rad. This is close to the observed uncertainties for our specific setup as shown in the previous section.\\
\\
In a first step, for each of the CTVs a set of PTVs with uniform margins of different margin size was automatically generated using TPS provided functionality. The margin sizes considered were specifically 0.05 cm, 0.1 cm, 0.2 cm, 0.3 cm, 0.4 cm, 0.5 cm and 0.6 cm around the CTV. Next, sets of anisotropic PTV were generated for different occurrence thresholds, for both CTVs and both uncertainty distributions considered. For each generated target a RT plan was generated as described in section~\ref{sec:planning}. For each plan, respectively for its resulting dose distribution the TCP was simulated in a Monte Carlo procedure as described in section~\ref{sec:tcpsim}.  This was done for a $10\times 4.0$ Gy fractionation schedule, which is how the patient was originally treated, as well as for a single fraction of $1 \times 17.023$ Gy. The single fraction dose was calculated to radio-biologically match the ten fractions treatment. In Fig.~\ref{fig:tcp_vol_comp}, for both CTVs the $TCP_{95\%}$ is plotted versus the excess volume for $\sigma_\theta=\sigma_\phi=0.1$ rad. The excess volume $V_{\text{excess}}$ stands for the additional volume which is added to the CTV volume by the particular PTV target. 
\begin{equation}
V_{\text{excess}}=V_{PTV}-V_{CTV}
\end{equation}
Figure~\ref{fig:art_ctv_0.02rad} is the same depiction for $\sigma_\theta=\sigma_\phi=0.02$ rad.
In the plots also the theoretical maximum TCP value is shown as a horizontal line. That is the TCP which would result if each voxel of the CTV would always be exposed to the prescribed dose.

\subsection{DVH and NTCP}
\begin{figure}
	\begin{subfigure}[b]{\textwidth}
         \centering
         \includegraphics[width=\textwidth]{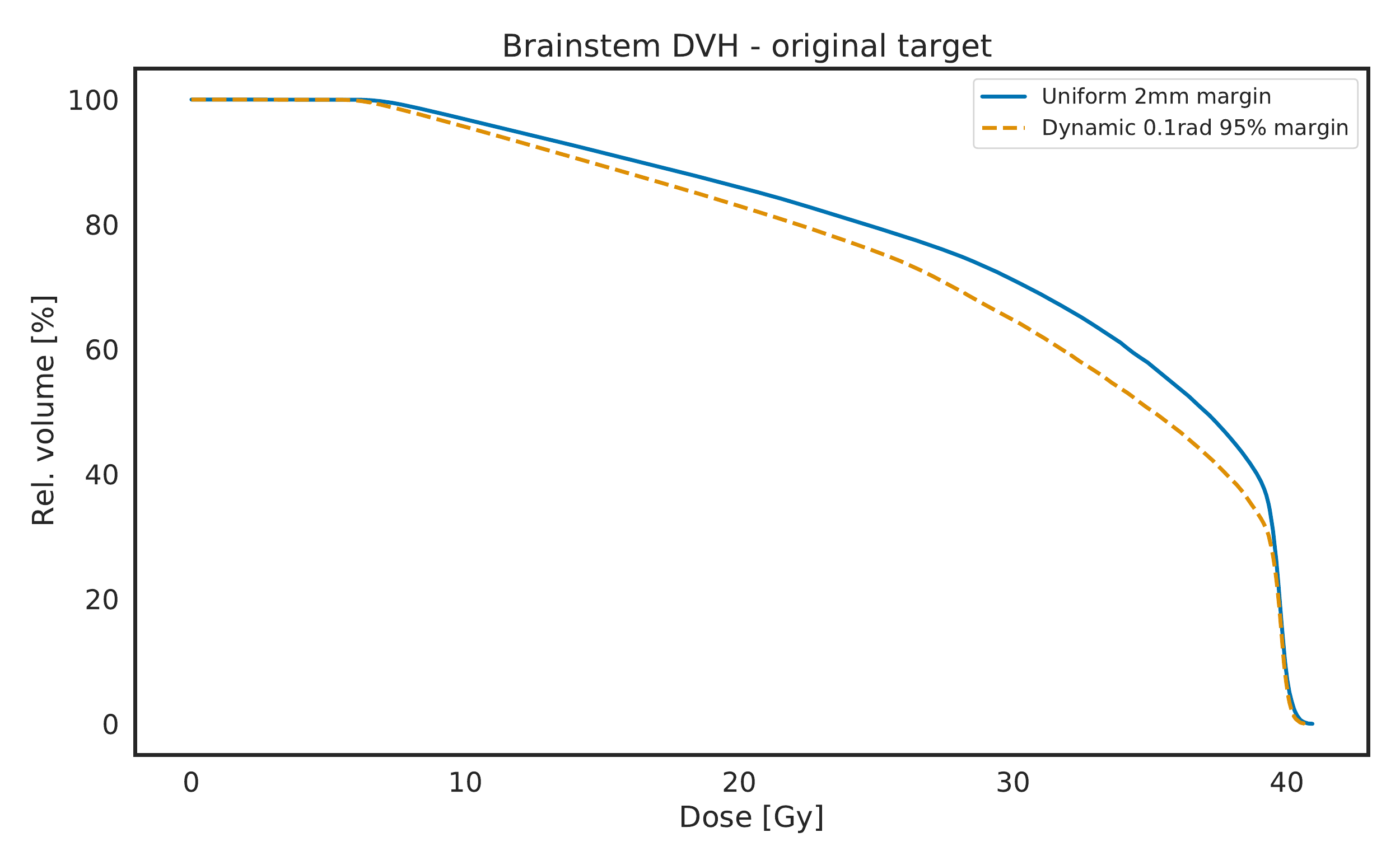}
         \caption*{A}
         \label{fig:org_dvh_0.1rad}
     \end{subfigure}
     \begin{subfigure}[b]{\textwidth}
         \centering
         \includegraphics[width=\textwidth]{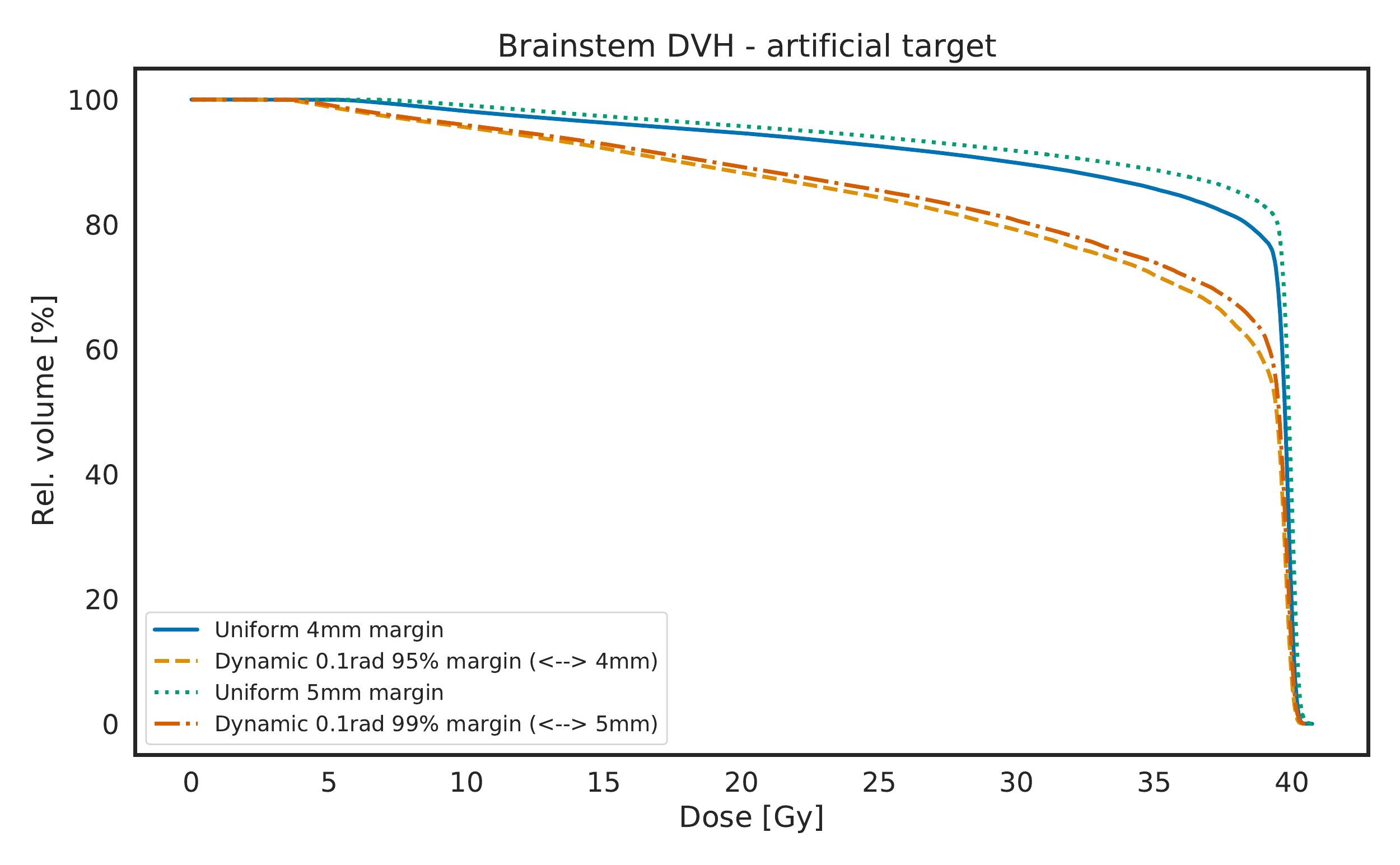}
         \caption*{B}
         \label{fig:art_dvh_0.1rad}
     \end{subfigure}
\caption{The plots show the DVHs of the brainstem for RT plans on uniform and corresponding anisotropic margin targets which yield the approximately same $TCP_{95\%}$ ; A: original CTV; B: artificial CTV}
\label{fig:dvh}
\end{figure}

To demonstrate the benefit of smaller target excess volume on the dose and NTCP of the OAR, which in our case is the brainstem, we plotted its cumulative dose volume histograms (DVH) (in Fig.~\ref{fig:dvh}) as well as computed and compared its NTCP. 
This was done for a uniform margin target and an anisotropic margin target which have approximately the same $TCP_{95\%}$. Looking at the plots in Fig.~\ref{fig:tcp_vol_comp} it can be observed that for the original CTV, such a case is given for the uniform 2mm margin target and the anisotropic target with $95\%$ threshold, which result in a brainstem NTCP of $12.4\%$ and $10.2\%$ respectively. For the artificial CTV, this is the case for the 5 mm uniform and $99\%$ threshold anisotropic targets, yielding brainstem NTCPs of $24.5\%$ and $18.2\%$ as well as for the 4 mm uniform and $95\%$ anisotropic targets, which yield $22.7\%$ and $17.2\%$.  For the NTCP calculation parameters from \citep{BURMAN1991123} were used.
The plan template used was created with the intent to ensure homogeneous dose coverage of the target and optimal conformity. Sparing of normal tissue or critical organs was not an objective, thus the absolute NTCP values are not relevant.

\section{Discussion and Conclusion}
In previous research \citep{Stroom1999, VanHerk2000, VanHerk2002,RotheArnesen2008,Deveau2010,Chen2011,Jin2011,Witte2017,Miao2019,Selvaraj2013}, the impact of setup uncertainties on the outcome of radiotherapy plans, and the consequences for appropriate margin derivation have been studied in various ways. However, most of this research work has certain limitations due to its methodology, such as considering artificial examples (spheres instead of real patient targets, artificial dose distributions instead of realistic RT plans), using DVH data which is subject to loss of spatial dose distribution information, or other shortcomings.  In this study we used a state of the art computational methodology in a novel approach to directly investigate the impact of rotational uncertainties on the tumor control probability depending on the CTV - PTV margin. A new anisotropic CTV to PTV generation algorithm was devised and implemented. The uniform and anisotropic margins were systematically compared for their impact on the resulting target volume and TCP. Specifically analysed was the correlation of the excess volume, with the TCP which is at least achieved in $95\%$ of the simulated treatments. The excess volume refers to the volume which a target adds to the primary CTV.  We presume that the excess volume can be, within certain limits, regarded as proportional to the damage inflicted to normal tissue. Fig.~\ref{fig:dvh} shows that the irradiated volume of the OAR is reduced and therewith the complication probability is smaller. Thus establishing a margin with a small excess volume which achieves the maximum possible TCP is desirable. \\
\\
The absolute values of the rotational setup uncertainties occurring in our particular setup were found to be very small.  Analysis of the histograms show that roll and pitch angles conform to a skew normal distribution around a mean angle which is very close to zero. This supports our hypothesis to assume the rotational uncertainties to be normally distributed around a mean equal zero. The currently used PTV generation procedure which consists of adding uniform margins of 1 - 2 mm around the CTV is adequate to compensate the observed rotational uncertainties. Nonetheless it is difficult to asses if further reduction of the margin is advisable due to other sources of uncertainty. A difficulty faced in the process of this investigation was the limited precision of the TPS environment, which is limited by factors such as CT slice thickness, the maximal precision of the optimization and dose calculation algorithms and other factors. The maximum possible precision of these parameters was in the same order of magnitude as the uniform margins of 1-2 mm currently used. This is especially of importance when the CTV volumes in question are mostly small volume structures, as the canine brain tumors considered in this study. 


In setups without anaesthesia, which do not allow for such rigid positioning, such as in human patients, noticeably larger rotational uncertainties are to be expected. In that case the anisotropic PTV was shown to be clearly superior to the uniform margin PTV recipe. It outperforms the uniform margin PTV by providing a larger TCP with a smaller excess volume.
\\
As expected, the geometrical shape of the target and its position relatively to the position of the ROSMA was shown to have a major influence on the resulting TCP - margin dependence. For the artificial target with cylindrical shape the rotational uncertainties have a significantly larger impact on the TCP. Furthermore, the benefit of the anisotropic margin targets over the uniform margin targets is greater. 

Another crucial aspect of this investigation is the fractionation. For the single fraction case, stochastically occurring setup uncertainties become systematic uncertainties. Fractionated treatment on the other hand can to some extent compensate for the stochastic uncertainties. This was also observed in the present study. 

The stochastic CTV to PTV generation method sometimes leads to artifacts in the generated target, which require manual postprocessing or repetition of the generation process. The improvement and refinement of the procedure could be the subject of future work.

A shortcoming of the treatment simulation is that in reality the rotation would apply to the entire body of the patient, while in the simulation only the CTV is rotated inside the static dose distribution. Thus, the dose distribution in reality might be minimally different from the static dose distribution calculated by the TPS due to the slightly shifted passage of the beams. We believe it is reasonable to assume that this discrepancy is negligible for the rotational uncertainties of magnitudes investigated in this study.

The methodology and the tools established for conducting this study can be easily extended to also investigate translational uncertainties and/or to simulate other relevant aspects such as inhomogeneous clonogenic cell density distribution within the CTV. This could be a potential subject of future research work.

\section*{Acknowledgment}
This work was supported by the Swiss National Science Foundation (SNSF), grant number: 320030-182490; PI: Carla Rohrer Bley

\clearpage
\appendix
\renewcommand\thefigure{\thesection.\arabic{figure}}

\counterwithin{figure}{section}
\section*{Appendix}
\setcounter{figure}{0}
\subsection{Application of the PTV expansion algorithm to spinal axis irradiation in human patients}

To demonstrate the benefits of the PTV expansion algorithm, we used it to establish a PTV margin for spinal axis irradiation. Due to large length of the spinal cord, its irradiation requires the use of multiple fields with different isocenters. Usually the setup is performed at the most cranial isocenter. As the different treatment fields are matched or overlapping, a compensation of rotational setup differences is not possible for the other isocenters. This can lead to rotational setup uncertainties, in particular around the yaw axis, which need to be compensated by an adequate margin around the CTV. Usually, in clinical routine, an isotropic margin is drawn around the CTV. Here we have tried to define an anisotropic margin which results from the rotational yaw uncertainties.\\
\\ For the PTV generation we assumed that the rotation point lies at the cranial end of the spinal coord (around $z=0$). We generated the PTVs for rotational variations around the yaw axis with $\sigma_\phi=$0.01 rad and $\sigma_\phi=$ 0.02 rad. We assume that these magnitudes approximately match the variations encountered in clinical reality. Figure~\ref{fig:bev} shows the beam eye view of the whole spinal axis. The purple contour delineates the CTV. The dark blue and the yellow contours depict the PTVs for $\sigma_\phi=$0.01 rad for thresholds of $95\%$ and $98\%$ respectively. The red and the black contours are the PTV for $\sigma_\phi=$0.02 rad for thresholds of $95\%$ and $98\%$ respectively. Figure~\ref{fig:transverse} shows the transversal view at the bottom end of the spinal cord (far away from the rotation point) with its latero-lateral margin expansion.

\renewcommand{\thefigure}{A\arabic{figure}}
\setcounter{figure}{0}


\begin{figure}
\includegraphics[width=\textwidth]{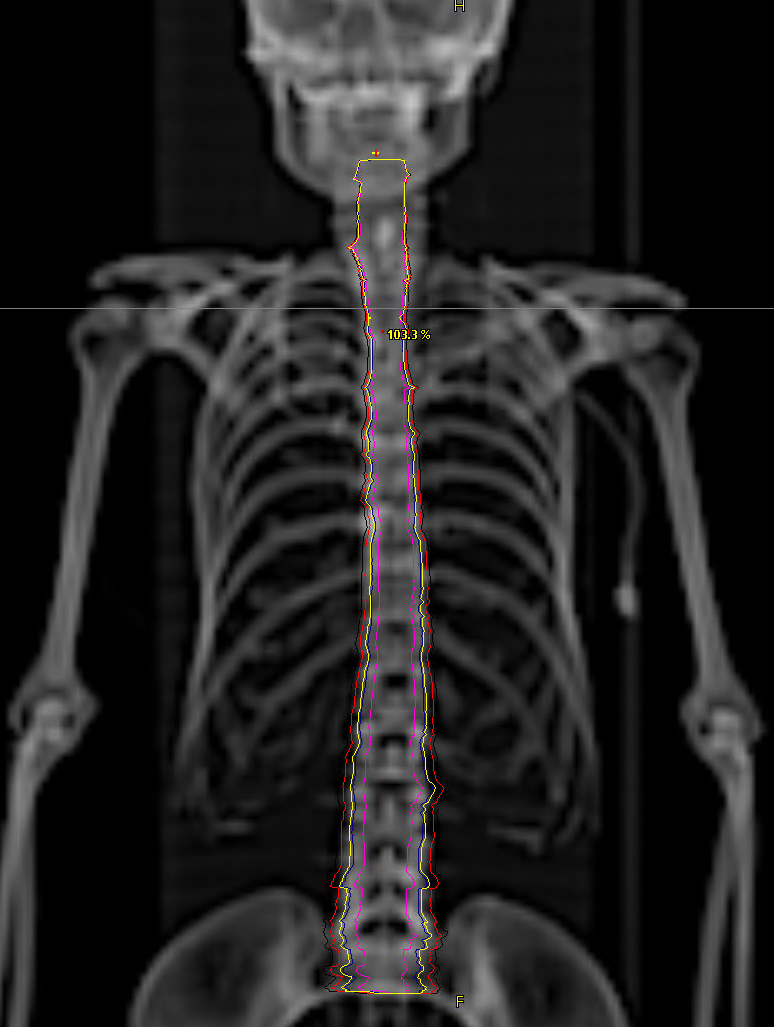}
\caption{Beam eye view of the spinal cord CTV and generated PTVs at the bottom end of the spinal cord. The dark blue and the yellow contours are the PTV for $\sigma_\phi=$0.01 rad for thresholds of $95\%$ and $98\%$ respectively. The red and the black contours are the PTV for $\sigma_\phi=$0.02 rad for thresholds of $95\%$ and $98\%$ respectively. }
\label{fig:bev}
\end{figure}

\begin{figure}
\includegraphics[width=\textwidth]{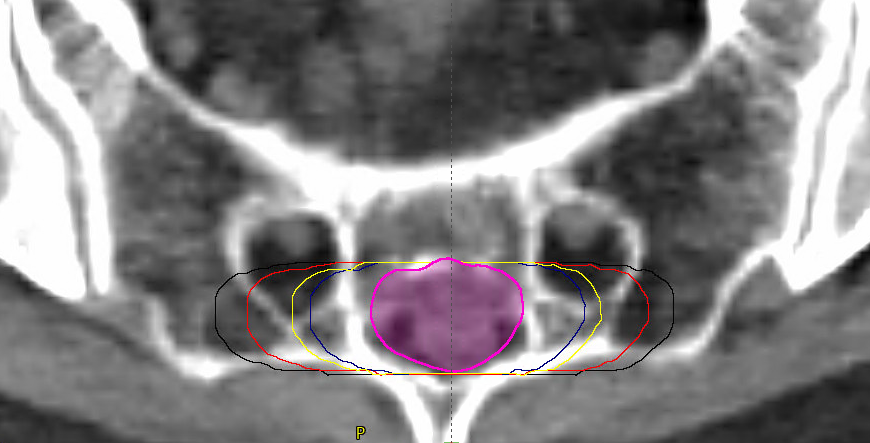}
\caption{Transverse view of the spinal cord CTV and generated PTVs at the bottom end of the spinal cord. The dark blue and the yellow contours are the PTV for $\sigma_\phi=$0.01 rad for thresholds of $95\%$ and $98\%$ respectively. The red and the black contours are the PTV for $\sigma_\phi=$0.02 rad for thresholds of $95\%$ and $98\%$, respectively. }
\label{fig:transverse}
\end{figure}

\begin{figure}
\includegraphics[width=\textwidth]{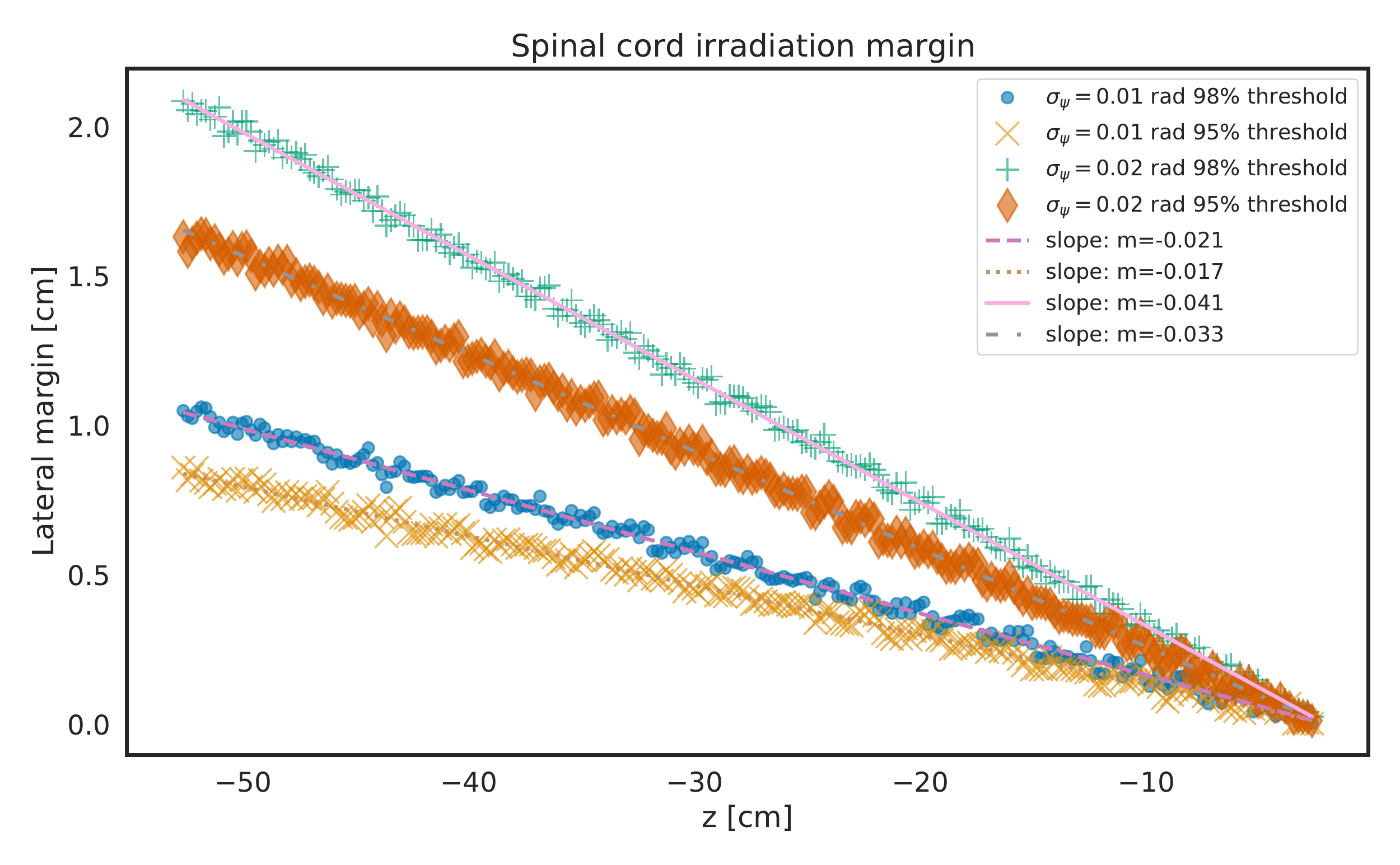}
\caption{Lateral margin between generated PTVs and the spinal cord CTV dependent on the distance from cranial isocenter $z$ }
\label{fig:margins_spinal}
\end{figure}

In Figure~\ref{fig:margins_spinal} we have plotted the lateral margin dependent on the distance from the cranial isocenter $z$. It can be observed that the margin is a linear function of the $z$. This allows a simple procedure for dynamic margin definition for spinal axis irradiation in clinical routine. In the plane of the most cranial isocenter a (small) margin which accounts only for the non-rotational uncertainties is drawn. At the most caudal plane of the spinal cord the lateral margin required is calculated and drawn using the slope. Then a simple interpolation is performed in between. This functionality is provided by the TPS. As an example: the lateral expansion margin needed at a distance of 40 cm from the cranial isocenter is 1.5 cm (assuming $\sigma_\psi = 0.01$ rad and $95\%$ threshold).

\clearpage

\bibliographystyle{unsrtnat}
\bibliography{literatur} 

\end{document}